\documentclass[twocolumn,epsfig]{revtex4}
\usepackage{epsfig,color}
\begin{document}
\title{Spectral ratio: an observable to determine $K^{+}$ nucleus potential and $K^{+}$ N scattering cross section}

\author{Aman D. Sood$^1$ }
\email{amandsood@gmail.com}
\author{Ch. Hartnack$^1$ }
\author{J\"org Aichelin$^1$ }
\address{
$^1$SUBATECH,
Laboratoire de Physique Subatomique et des
Technologies Associ\'ees \\University of Nantes - IN2P3/CNRS - Ecole des Mines
de Nantes 
4 rue Alfred Kastler, F-44072 Nantes, Cedex 03, France}
\date{\today}

\maketitle

\section*{Introduction}
The change of the properties of mesons in dense hadronic matter
has theoretically been investigated since many years
\cite{Aichelin:1986ss,hartphysrep} and an experimental verification is still
missing. The $t\rho$ approximation allows predicting the optical
potential of the mesons in low-densities matter by experimentally
measured phase shifts. At higher densities more sophisticated
approaches have to be employed and in the last two decades many
efforts have been made to investigate the properties of $\rho$,
$\omega$, $K^{+}$ and $K^{-}$ mesons in matter.
These calculations are complex  because most of the mesons can
form baryonic resonances which have other decay branches.
Therefore, coupled-channels calculations have to be employed and
the challenge has been met to calculate them self-consistently.
Nevertheless, the theoretical predictions launched by different
groups differ substantially because several of the quantities
which enter such calculations, like in-medium coupling constants
and the in-medium dressing of the different particles are only
poorly known. These uncertainties render the theoretical
prediction rather vague and it is highly desirably to have
experimental information of properties of mesons around and above
normal nuclear matter densities.\
Here we aim to show that the ratio of the momentum spectra of $K^{+}$ at small transverse
momentum measured for symmetric systems of different sizes can be such an observable. For the present study we use IQMD model. For details please refer to \cite{hartphysrep}
\begin{figure}[!t] \centering
\vskip 0.5cm
\includegraphics[angle=0,width=6cm]{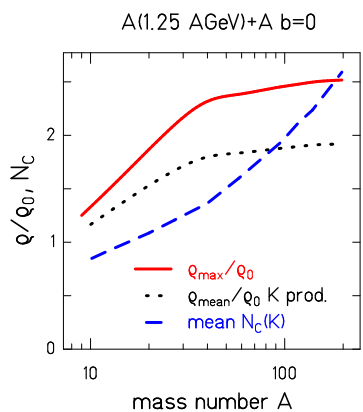}
\caption{\label{fig1} Density and maximum number of collisions of $K^{+}$ as a function of mass of the system. Various lines are explained in the text.}
\end{figure}
\section*{Results and Discussion}In fig 1, we display the
maximum value of the central density (full line) as a function of
the projectile mass number $A$ in symmetric reactions $A+A$ at
1.25 AGeV incident energy. The density rises strongly up to about
$A=40$ and then only slightly for the higher masses. A similar
conclusion can be drawn when looking on the mean density at the production
points of kaons (dotted line). These densities
enter directly into the $K^{+}$ nucleus potential and have an
important influence on the production yield of the kaons. Again,
the yield rises up to $A=40$ and then saturates at higher masses.
In contrast, the mean number of collisions $N_C(\rm K)$ which a
kaon suffers before leaving the system (dashed line), rises quite
moderately with $A$ up to $A=40$ and then increases much stronger
for heavier systems. Therefore, the large mass region is the realm
for measuring $ \sigma^{\rm{medium}}(\rm{K^+N}\to \rm{K^+N})$.
Below projectiles of mass 40 we see a strong increase of the
densities with system size but few rescattering collisions because
the systems are so small that rescattering does not become
important. This is the realm for measuring  $K^{+}$ nucleus potential.

\begin{figure}[!t] \centering
\vskip 0.5cm
\includegraphics[angle=0,width=6cm]{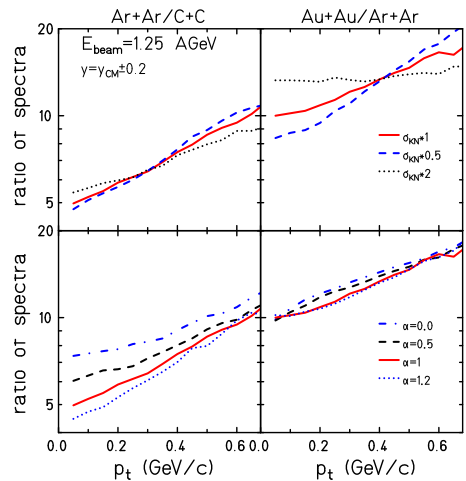}
\caption{\label{fig2}Ratio of
transverse momentum spectra for symmetric systems of different
sizes. Various lines are explained in the text.}
\end{figure}

Figure~\ref{fig2} shows the preliminary results of our calculations. It
presents the ratio of the transverse momentum spectra close to
midrapidity, obtained in Ar+Ar and C+C collisions (left) and Au+Au
and Ar+Ar collisions (right). The top panels show the influence of
the rescattering cross section. The free KN rescattering cross
section has been multiplied by a coefficient between 0.5 and 2
while leaving all the other parameter unchanged. The bottom panel
displays the variation of this ratio with the change of the
strength of the $K^{+}$ nucleus potential by applying a factor
$\alpha$ to the potential.
As expected from the discussion above, the ratio of the yields for
the two smaller systems is almost independent of the rescattering
cross section but depends strongly on the strength of the $K^{+}$
nucleus potential. Rescattering is not very frequent in these
light systems, therefore its  influence on the spectrum is
moderate. The $K^{+}$ nucleus potential, in the contrary, has a direct
influence on the spectra, as  shown in the lower left panel. The
maximal density as well as the density profile is different for
the both systems and due to this difference the $K^{+}$ nucleus
potential acts differently. Whereas in C+C even the central
density does not exceed much the normal nuclear matter density, in
the Ar+Ar system the densities exceeds already twice normal
nuclear matter density. By comparing the lighter systems one cannot learn
much on the rescattering cross section but the spectra becomes
sensitive to the strength of the $K^{+}$ nucleus potential. The slope
of the ratio is a direct measure of this strength and present-day
experiments are sufficiently precise to extract the strength of
the potential.

If one compares the two heavier systems (right panel of
Fig.~\ref{fig2}, a completely different scenario emerges.
Rescattering becomes very important in the Au+Au reactions where
almost all $K^{+}$ undergo rescattering. Therefore, the influence of
the rescattering cross section on the spectra is very visible as
shown in the upper right panel. On the other side, the ratio is
hardly influenced by the strength of the $K^{+}$ nucleus potential
because
the density at which the $K^{+}$ have their last collisional
interaction with the surrounding nucleons is rather similar and
therefore, the $K^{+}$ nucleus  potential acts in a very similar way
on the \mbox{K$^+$}, leaving the ratio unchanged.

\section*{Acknowledgments}
This work has been supported by a grant from Indo-French Centre for the Promotion of Advanced Research (IFCPAR) under project no 4104-1.


\end{document}